\documentclass[%
 aip,
rsi,%
 amsmath,amssymb,
 reprint,%
]{revtex4-1}

\usepackage{graphicx}
\usepackage{dcolumn}
\usepackage{bm}

\begin{document}

\preprint{AIP/123-QED}

\title[]{A Coalition of Single Slit and Double Slit Diffraction}

\author{Supantho Rakshit}
 
 \email{sukantoraxit@gmail.com}
 
\affiliation{Notre Dame College, Arambagh, Motijheel, Dhaka-1000, Bangladesh
}%

\date{\today}

\begin{abstract}
In this paper, we theoretically investigate a particular experimental setup which coalesces the concepts of the double slit and single slit diffraction. In Thomas Young's classic double slit experiment, monochromatic plane light wave impinges on an opaque screen with two parallel, long and narrow slits and the resulting intensity pattern is observed on a screen placed at a great distance placed from the slits. We show, via theoretical calculations that the if another opaque screen with a single long narrow slit is placed at a distance from the two slits, the resulting intensity pattern on the observation screen is nontrivial and interesting.
\end{abstract}

\maketitle

\section*{Introduction}
\label{intro}
Consider two classic topics: Thomas Young's famous double slit experiment and single slit diffraction. These two topics are one of the most widely taught
and demonstrated topics in any introductory courses in optics. These two experiments have been used as introductory examples of optical interference and
diffraction in many classic textbooks such as Jenkins and White \cite{white}, Born and Wolf \citep{wolf}, Hecht \citep{hecht} etc.\\
In the double slit experiment, monochromatic coherent plane light wave falls on an opaque screen with two narrow and long slits and the resulting interference
pattern is observed on a screen placed at distance from the slits. The overall setup in single slit diffraction is the same, except that the slit number here is
only one. The intensity pattern along the observation screen varies as cosine squared and $sinc$ squared in double slit and single slit diffraction respectively.
We set forth an inquiry; is there a possible way to combine the core concepts of these two topics? Can we develop an experimental setup that merges double
slit and single slit diffraction?\\
In this paper, we will propose an experimental setup that does so.
\section*{The Experimental Setup}
Monochromatic coherent plane light wave of wavelength $\lambda$ falls perpendicularly on an opaque wall with two narrow and long slits separated by a distance $d$.
From a distance $L_{1}$ from the slits, another opaque wall is placed with a single slit of angular width 8 degrees about point O. Then, an observation screen is
placed from a distance $L_{2}$ from the single slit. The overall setup in shown in Fig. 1.
\begin{figure}[h]
\label{fig1}
\includegraphics[width= 9cm, height=18.4cm,keepaspectratio]{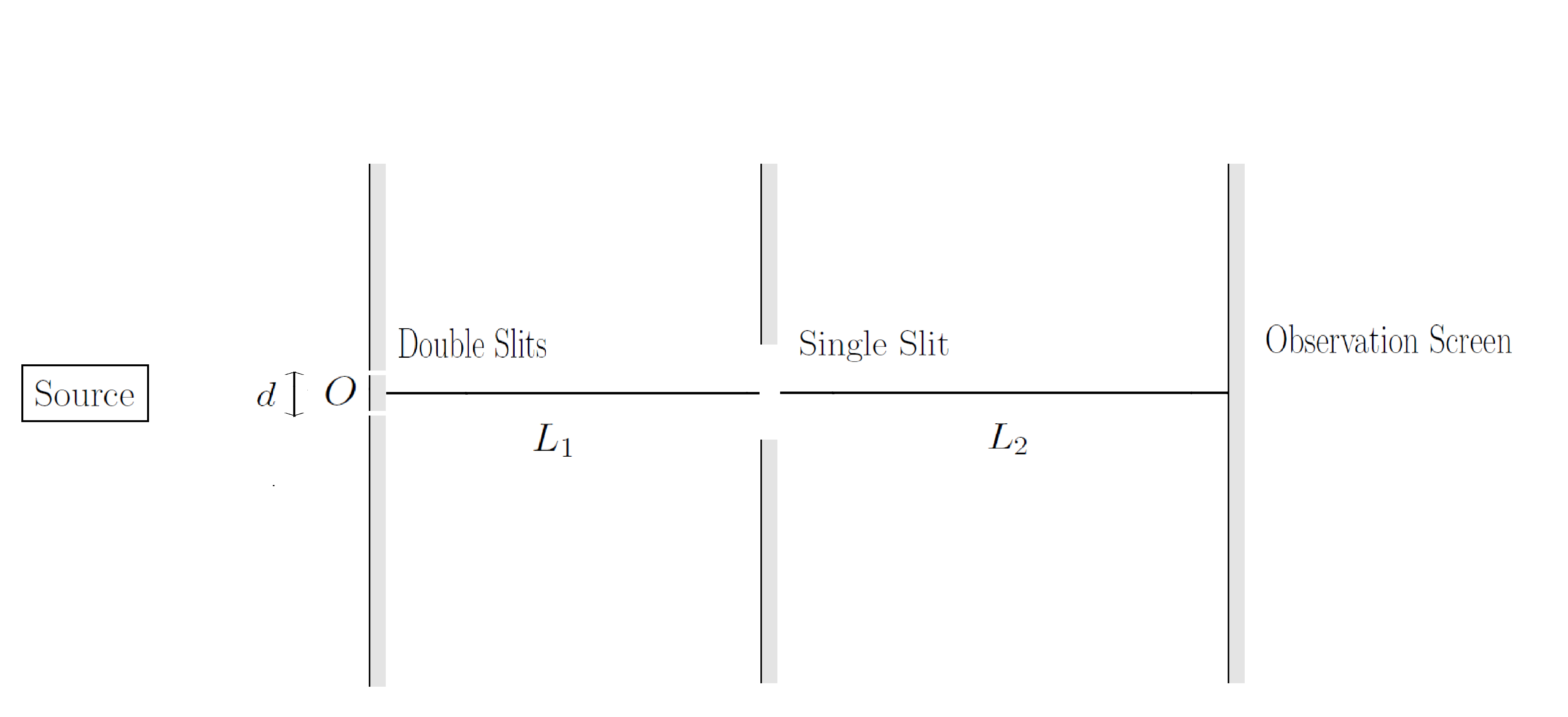}
\caption{Monochromatic coherent light waves impinge on two slits separated by a distance $d$ from a light source. A single slit of angular width of $8^ \circ$ about from the midpoint of the double slits $O$ is placed a distance $L_{1}$ from the double slits. Then an observation screen is placed a distance $L_{2}$ from the single slit, where the final interference pattern will be observed.}
\end{figure}

\section*{Finding The Intensity Distribution On The Observation Screen}
Here, at first the light from the source impinges on the two slits. All the points on the two slits act as Huygens sources of secondary wavelets and forward the disturbance incident upon them. Then the forwarded disturbance reaches the single slit placed a distance $L_{1}$ in front of the double slits. In order to find the resulting intensity distribution on the screen, we will assume that all points on the single slit act as a continuous array of virtual harmonic oscillators (or so called Huygens sources), which are the sources of secondary wavelets. We will add up all the contributions from these virtual oscillators using phasor addition.\\
See Fig 2.\begin{figure}[h]
\label{fig1}
\includegraphics[width= 10cm, height=19cm, keepaspectratio]{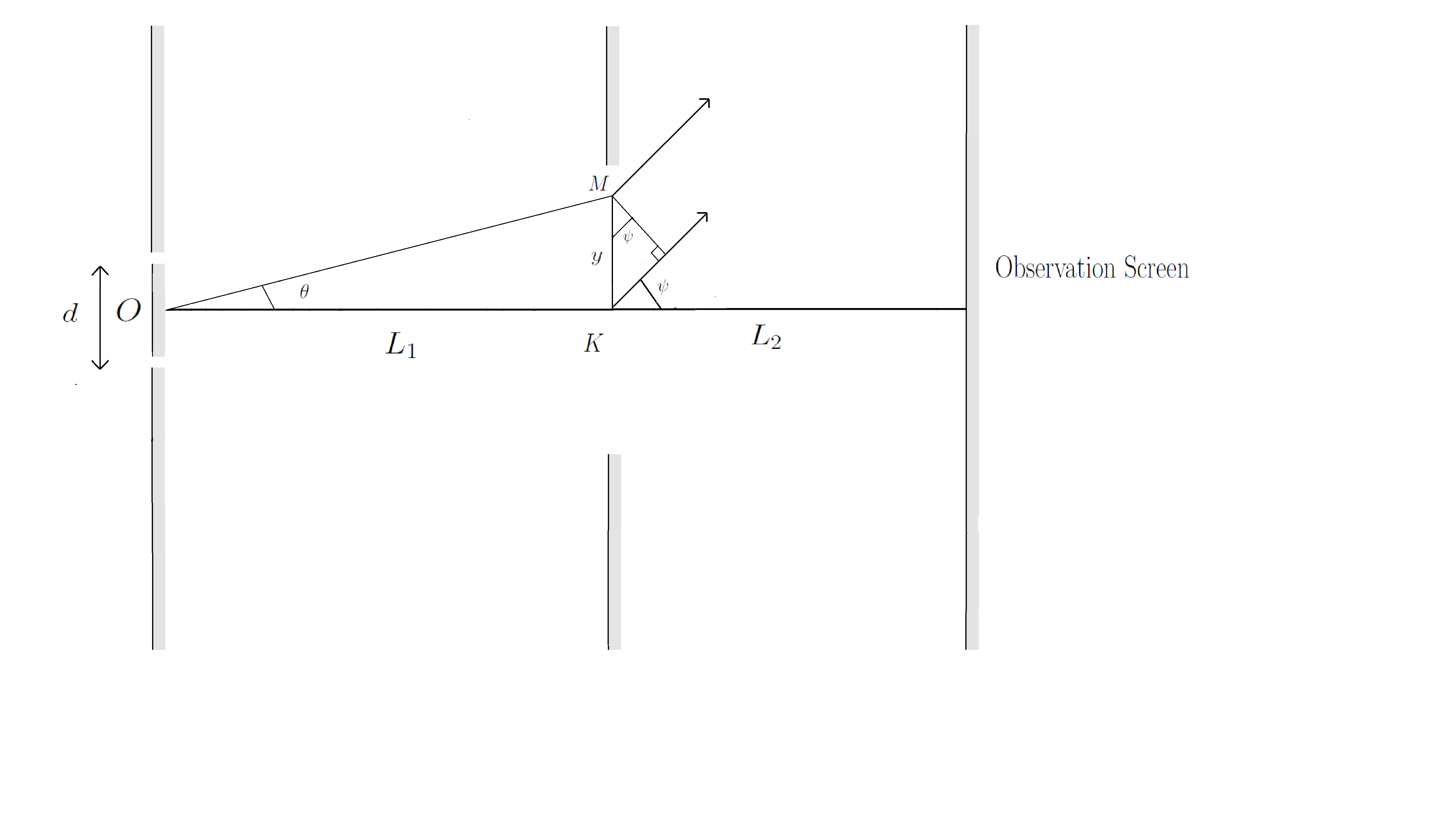}
\caption{All the points on the single slit act as a continuous array of secondary wavelets. Point $K$ is the midpoint of the single slit and point $M$ is situated at an angle $\theta$ from point $O$ and a distance $y$ from point $K$. For an observation angle $\psi$, there is a path difference of $y \sin \psi$ between $K$ and $M$. With respect to point $K$, there is also an additional path difference of $d \sin \theta$ for the disturbance emanating from point $M$ because the effective contribution from point $M$ started a time $(d \sin \theta)/c$ later than that of point $K$; due to a path difference of $d \sin \theta$ for the double slits fro angle $\theta$.}
\end{figure} Consider two points on the slit: point $K$ and $M$. Point $K$ is the midpoint of the slit and point $M$ is at a distance $y$ from $K$ and situated at an angle $\theta$ from point $O$.
From the Fig 2, we easily see that for an observation angle $\psi$ on the screen, there is a path difference of $y \sin \psi$ between points $K$ and $M$. But there is another issue here. Point $K$, being the midpoint of the single slit, received the disturbance from the two slits simultaneously, yielding a zero path difference between the incident disturbances for that point. But the virtual oscillator at point $M$ was effectively "switched on" after a time $d \sin \theta/c$ after point $K$ was "switched on"; this is due to the fact that for angle $\theta$, the light waves from the two slits had a path difference of $d \sin \theta$. So the "effective" contribution from point $M$ will start a time  $d \sin \theta/c$ later than that of point $K$.\\
Hence, the net path difference is
\begin{equation}
\label{eq1}
    \Delta x = d (\sin \psi - \sin \theta) = d \left(\frac{y_s}{L_2}-\frac{y}{L_1}\right).
\end{equation}
Here $y_s$ is the point on the observation screen corresponding to the observation angle $\psi$. We have used the small angle approximation for both $\psi$ and $\theta$.\\
Now, the contribution from point $M$ is proportional to
\begin{enumerate}
    \item the oscillator's infinitesimal length $dy$, due to size-source proportionality, 
    \item the net amplitude of the disturbance impinging on it, that is $\cos \left(\frac{\pi dy}{\lambda\ L_1}\right)$. The amplitude of the net disturbance from the double slits varies along the $y$ cordinate in this manner \cite{hrw},
    \item $e^{i\ \frac{2\pi }{\lambda} \Delta x}$, taking account the phase difference.
\end{enumerate}
So, the overall contribution from $M$ is proportional to \begin{equation}
\label{eq2}
\cos \left(\frac{\pi dy}{\lambda\ L_1}\right) e^{i\ \frac{2\pi }{\lambda} \Delta x} dy .
\end{equation}
 The total field can be calculated by integrating over this expression.\\
Now, from symmetry, we see that the total contribution from the top half and the bottom half of the single slit are the same. In order to find the total field at an observation angle $\psi$ (from point $K$ to the observation screen), we will sum up the contributions from the two halves, taking account a path difference of $L_{1}\ 4^{\circ} \approx 0.07 L_{1}$ between them.\\
In order to find the overall contribution from a single half, we have to integrate over Eq.\ref{eq2} from $y=0$ to $y=L_{1} \sin 4^{\circ} \approx 0.07 L_{1}$.
Here $d, \lambda, L_{2}$ and $L_{1}$ are all constants. The result of the integration will be a complex quantity. The amplitude of the derived complex quantity will be the amplitude of the total disturbance emanating from half the slit. So, performing the integration and calculating the amplitude, we find that the amplitude of the total disturbance from half the slit is
\begin{equation}
E_{half}= \xi \frac{L_1 \lambda\ }{d}\left\{\sqrt{1+\cos ^2\left(\frac{.07\pi d}{\lambda}\right)}-2\right\}.
\end{equation}
Here. $ \xi $ is a constant. We see that $ L_{1}, \lambda, d$ are all constants and there is no $y_{s}$ dependence. Hence $E_{half}$ is a constant itself too.\\
Now there is a path difference of $L_{1} 4^\circ \sin \psi \approx 0.07L_{1} \sin \psi$ between the contributions the two halves of the slits for observation angle $\psi$. Since they have equal effective amplitude of $E_{half}$, the final intensity distribution on the screen will be equal to
\begin{equation}
I_s = 4\ E_{half}^2\ \cos ^2\left(\frac{0.07 \pi L_{1} y_s}{\lambda \ L_2}\right).
\end{equation}
Where we have used the small angle approximation $\sin \psi \approx \tan \psi=y_{s}/L_{2}$.\\
 This means the intensity along the observation screen varies as cosine squared. Therefore, the intensity pattern on the screen is similar to a two beam interference, in contrast to the $sinc$ squared type pattern usually observed for single slits. The reason is that the two halves of the slits acted as two composite extended, yet coherent source of disturbance. Another interesting thing is that, if the single slit were not placed, the positions of maximas on the screen would have been given by
\begin{equation}
y_{s} = m \frac{\left(L_{1} + L_{2} \right) \lambda}{d}.
\end{equation} 
And from Eq. $4$, it can be deduced that the positions of maximas on the screen for the described setup is
\begin{equation}
y_s= m\ \frac{\lambda L_2}{0.07\ L_1}.
\end{equation}
Where in both cases $m$ is an integer. So we can easily see that there is a phase shift between the two cases. The fringe shift is equal to
\begin{equation}
\label{eq7}
\Gamma = \frac{y_s}{\lambda}\left(\frac{0.07L_1}{L_2}-\frac{d}{L_1+L_2}\right).
\end{equation} 
Here we see that the fringe shift has a linear dependence on $y_{s}$. Using this property, the quantities $d, L_1, L_2$ or $ \lambda$ can be determined with high accuracy  Eq, \ref{eq7}. We also can deduce from Eq. $5$ and Eq. $6$ that the width of the maximas (or minima) has changed from $(L_1 + L_2) \lambda /d$ to $\lambda L_2/ 0.07 L_1$. The changed in width is
\begin{equation}
\delta = \lambda \ \left(\frac{L_1+L_2}{d}-\frac{L_2}{0.07\ L_1}\right)
\end{equation}
Eq. $8$ can also be helpful for precise calculation of $d, L_1, L_2$ or $ \lambda$.
\section*{Conclusion}
We have proposed an experimental setup which fuses the single slit and double slit diffraction. The setup was similar to Thomas Young's classic double slit interference. We modified it by placing an opaque screen with a narrow long slit before the observation screen, at a certain distance from the double slits and theoretically found that the resulting intensity pattern completely dissimilar to that of the normal single slit; rather its is analogous to a two beam interference pattern with a fringe shift from the previous case of double slit interference pattern. We have also proposed two possible methods: (i) fringe shift and (ii) change in the width of maxima/minima; for calculating the the distance between double slits, the wavelength of light or the opaque screen distances.  This topic can be of high interest in a physics classroom during an introductory course of optical interference. This may also tempt the students to investigate and explore various aspects of other diverse optical experiment setup. 

\end{document}